\documentclass[twocolumn,eqsecnum,showpacs,pra,superscriptaddress]{revtex4}
\usepackage{amsmath}
\usepackage{amssymb}
\begin{document}

\title{Continuous-variable teleportation  in the characteristic-function 
description }
\author{Paulina Marian} 
\affiliation{ Department of Chemistry, 
University of Bucharest, Boulevard Regina Elisabeta 4-12, 
R-030018 Bucharest, Romania}
\author{Tudor A. Marian}
\affiliation{ Department of Physics, 
University of Bucharest, P.O.Box MG-11,
R-077125 Bucharest-M\u{a}gurele, Romania}
\date{\today}

\begin{abstract}
We give a description of the continuous-variable
 teleportation protocol  in terms 
of the characteristic functions of the quantum states involved.
The Braunstein--Kimble protocol is written for an unbalanced
homodyne measurement and  arbitrary input and resource states. We 
show that the output of the protocol is a superposition between the
 input one-mode field and a classical one induced by measurement and 
classical communication. 
We choose to describe the input state distortion through  teleportation 
by the average photon number of the measurement-induced field. Only in the case of symmetric Gaussian resource states we find a relation between the optimal added noise and the minimal EPR correlations used to define inseparability.

\end{abstract} 
\maketitle

\section{Introduction}
The principal idea
of the teleportation process first described in Ref.\cite{Bennett} 
is that two distant operators, Alice at a sending station and Bob 
at a receiving terminal, share an entangled  bipartite quantum state 
and exploit its {\em nonlocal} character as a quantum resource.
In the continuous-variable case, the resource state which
 is also called an Einstein-Podolsky-Rosen 
(EPR) state \cite{EPR} is a two-mode entangled quantum state. 
The protocol for teleportation of quantum radiation field states originally 
proposed by
 Braunstein and Kimble \cite{BK} within the Wigner-function formalism
was subsequently given in several other descriptions. Wave-function
treatments in the Fock-state basis \cite{Enk}, continuous-coordinate 
representation \cite{MB,Opat} or in the coherent-state expansion
\cite{Vuk} are particularly useful in the pure-state input case 
and ideally faithful
 teleportation. Note also the formalism of the transfer operator
 put forward in Ref.\cite{H1}. Nonideal teleportation was 
treated by considering either a mixed entangled two-mode state 
\cite{OL}
shared by the two distant partners Alice and Bob or imperfect Bell measurements 
\cite{BK,Vuk}. Phase-space descriptions were given using the
$P$-representation of the density operator
 in a treatment valid also for mixed-state input \cite{Ban1}.
The  teleportation fidelity of mixed  one-mode nonclassical Gaussian states 
through a symmetric channel was quantified in our work Ref.\cite{PTH03}
by using the Uhlmann fidelity. A relation between 
the entanglement of the symmetric resource state 
measured by Bures distance \cite{PTH03b} and the accuracy of 
teleportation was then analyzed.
The  Wigner function  was used in 
describing teleportation as a general conditional measurement
 \cite{Paris} or to extend the protocol to non-Gaussian states  in Ref.\cite{CW}.
A lot of work was paid to analyze realistic ideas of 
increasing the teleportation fidelity. A recent survey on 
these issues is the e-print   
\cite{PM}. The role of teleportation in the continuous-variable 
quantum information is analyzed in the review Ref.\cite{BL}.
Especially related to our present work  are several 
papers investigating the possibility of increasing the quality of teleportation
by local interventions of the two operators \cite{F,MF,AI}.By changing the protocol, in Ref.\cite{HK},  the 
quantum nondemolition interaction is used to obtain entangled states for
 teleportation. There the incoming field and the 
two-mode resource field  are superposed at an unbalanced beam splitter
and the teleportation accuracy is discussed in terms of added noise to the
input field.
Note finally that in the recent e-print \cite{WGG}
the teleportation protocols were efficiently included in a more general
evaluation of the capacity of Bosonic channels.

In the present paper we give a description of the continuous-variable
 teleportation protocol \cite{BK} in terms 
of the characteristic functions (CF's) of the quantum states involved.
Our main tool in deriving the CF of the teleported state is 
the Weyl expansion of the density operator.
In Sec. II we consider the Braunstein--Kimble protocol written for an unbalanced
homodyne measurement and  arbitrary input and resource states. We 
show that the CF of the teleported state is the product between the
 CF of the input state and a function only depending on the 
properties of the two-mode resource state and the geometry of measurement.
 We identify 
this function as the normally ordered CF of a measurement-induced 
field. In Sec. III we concentrate on the teleportation of an arbitrary one-mode 
state
through a Gaussian channel and discuss the properties of
 the measurement--superposed field state. 
We show that this is a classical state, 
namely it possesses a well defined $P$-representation.
We choose to describe the input state distortion through  teleportation 
by the average photon number of the measurement-created field.
In Sec.IV we analyze the possibility of minimizing this variable by local 
interventions on the resource state. We conclude by elaborating on
the recently found relation   between optimal fidelity
of teleportation and the entanglement of the resource state \cite{AI}.

\section{Unbalanced homodyne measurement}
The density operator of the input one-mode  state $\rho_{in}$ 
(to be teleported) and that of
 the two-mode state $\rho_{AB}$ shared by Alice and Bob
 can be written as Weyl expansions
\begin{eqnarray}
\rho_{in}=\frac{1}{\pi}\int {\rm d}^2 \lambda \;\;\chi_{in} (\lambda)
D(-\lambda),\label{cf1}
\end{eqnarray}
and
\begin{eqnarray}
\rho_{AB}&=&\frac{1}{\pi^2}\int {\rm d}^2 \lambda_1 {\rm d}^2 \lambda_2\;\;
\chi_{AB}(\lambda_1,\lambda_2)D_1(-\lambda_1)D_2(-\lambda_2),\nonumber \\&&
\label{cf2}
\end{eqnarray}
where  $D(\alpha)=\exp{(\alpha a^{\dag}-\alpha^* a)}$ is a Weyl 
displacement operator and $a$ denotes the annihilation operator. 
Generally, the Weyl expansion of a Hilbert-Schmidt operator 
such as the density operator is a powerful tool in quantum optics and
points out the well known one-to-one correspondence between
 the density operator of a $ n$-mode field state $\rho$
and its  CF
$\chi(\lambda_1,\lambda_2\cdots \lambda_n):={\rm Tr}[\rho D(\lambda_1)
D(\lambda_2)\cdots D(\lambda_n)]$. 
We consider that initially the overall state
 of the system is the three-mode product state 
$\rho_{in}\otimes\rho_{AB}$ and adopt the 
philosophy of the original teleportation idea \cite{Bennett}:
 the input state $in$ remains unknown to Alice. For the moment
 we also assume that the states $\rho_{in}$ and $\rho_{AB}$ 
are arbitrary. In the first
 stage of the protocol, by using an unbalanced beam splitter 
with the rotation angle $\theta$ Alice measures simultaneously the variables 
\begin{equation} \hat q=\cos \theta \hat q_{in}-\sin \theta \hat q_A,\;\;
 \hat p=\sin \theta\hat p_{in}+\cos \theta \hat p_A, \label{b1} \end{equation}
where 
\begin{equation}\hat q_j=(\hat a_j+\hat a_j^{\dag})/{\sqrt{2}},\;\;
 \hat p_j=(\hat a_j
-\hat a_j^{\dag})/(\sqrt{2}i) \label{op}\end{equation}
are the canonical operators and $\hat a_j $ and $\hat a_j^{\dag}$,  
$j=1,2$, are the amplitude operators 
of the modes. A common eigenfunction 
of the pair of continuous
 quantum variables 
$\{\hat q, \hat p\}$ can be best written in the coordinate representation as
the continuous expansion \cite{L}
\begin{eqnarray} |\Phi(q,p)\rangle&=&\frac{1}{\sqrt{2\pi}\cos\theta }
\int_{-\infty}^{\infty} 
{\rm d}\xi\exp{(i p\xi/{\cos \theta})}
|\xi\rangle_A\nonumber\\&& \otimes|\xi 
\tan \theta+\frac{q}{\cos \theta}\rangle_{in},\label{a10}\end{eqnarray}
 where $(q,p)$ is the outcome of the homodyne measurement. The 
distribution function of the outcomes predicted by quantum mechanics is
\begin{eqnarray}{\cal P}(q,p)&=&{\rm Tr}_{in,AB}\left\{\left[|\Phi(q,p)\rangle
\langle\Phi(q,p)|\otimes I_B\right](\rho_{in}\otimes\rho_{AB})\right\}.
\nonumber\\&& 
\label{p}
\end{eqnarray}
As a consequence of the nonlocal character of quantum mechanics, 
the mode B at Bob's remote side is projected onto the state 
\begin{eqnarray}\rho_B(q,p)&\sim 
&
{\rm Tr}_{in,A}\left\{\left[|\Phi(q,p)\rangle
\langle\Phi(q,p)|\otimes I_B\right]
(\rho_{in}\otimes\rho_{AB})\right\}.\nonumber\\&& 
\label{B}
\end{eqnarray}
Our point here is to provide  a formula for the CF $\chi_{B}(\lambda_2,q,p)$ of 
the state \ (\ref{B}) via the Weyl expansions \ (\ref{cf1}) and \ (\ref{cf2}). 
To get this, we have first  derived the following  trace-formula
\begin{eqnarray}
{\rm Tr}\left[ |\xi\rangle\langle\xi^{\prime}|D(-\lambda)\right]&=&
\exp{[i\Im m{(\lambda)}(\Re e(\lambda)-\sqrt{2}\xi) ]}
\nonumber\\&& \times \delta\left[\xi^{\prime}-\xi+
\sqrt{2}\Re e(\lambda)\right],\label{a7}\end{eqnarray}
and applied it to the evaluation of the state \ (\ref{B}). Then,
  carefully performed Gaussian integrals led us to 
\begin{eqnarray}\chi_B(q,p,\lambda_2)&&=
\frac{1}{{\cal P}(q,p) \sin (2\theta)}\frac{1}{\pi^2}\nonumber\\&& \times
\int {\rm d}^2 \lambda \exp{(\mu \lambda^*-\mu^*\lambda)}  \;\chi_{in} (\lambda)
\nonumber\\ &&\times\chi_{AB} \left(\cot \theta \Re e{\lambda}-i \tan \theta \Im m{\lambda},\lambda_2
\right).\nonumber\\&&\label{B1}\end{eqnarray}
Under the integral above, the outcome $(q,p)$ of the measurement 
is just contained in the displacement parameter  
\begin{eqnarray}\mu=\frac{1}{\sqrt{2}}\left (\frac{q}{\cos \theta}+i \frac{p}{
\sin \theta}\right).
\label{mu}\end{eqnarray}
 Equation \ (\ref{B1})  suggests that Bob should perform a displacement
of his  state with the outcome parameter $\mu$. 
 The communication by classical means
 between Alice and Bob is thus an important step of the teleportation protocol. 
As a result of knowing 
 the displacement $\mu$  for every measurement, Bob can finally construct
the teleported state through an
ensemble  averaging with the distribution function 
\ (\ref{p}) over all measurement outcomes.
Note that with the well known property ${\rm Tr}_B D_B
(-\lambda_2)=\pi \delta^2 (\lambda_2)$ we can readily write 
the distribution function of the outcomes 
\begin{eqnarray}{\cal P}(q,p)&=& \frac{1}{\sin (2\theta)}\frac{1}{\pi^2}
\int {\rm d}^2 \lambda \exp{(\mu \lambda^*-\mu^*\lambda)}\chi_{in} 
(\lambda) \nonumber\\ 
&& \times 
\chi_{AB} \left(\cot \theta \Re e{\lambda}-i \tan \theta \Im m{\lambda},0
\right).\label{B2}\end{eqnarray}
The output density operator at Bob's side after the phase-space translation
\begin{eqnarray}
\rho_{out}&=&\int\int {\rm d}q{\rm d}p{\cal P}(q,p)D_B(\mu)\rho_B(q,p)
D_B(-\mu),\nonumber\\&&\label{out}\end{eqnarray}
has a one-mode Weyl expansion as in Eq.\ (\ref{cf1}).  The weight 
 function of this expansion is the CF of the teleported state 
which proves to have a remarkably factorized form
\begin{eqnarray}\chi_{B,out}(\lambda_2)&=&
 \chi_{in}(\lambda_2)
\nonumber\\&&\times
\chi_{AB}\left[\Re e(\lambda_2)\cot{\theta}-i \Im m ( \lambda_2) 
\tan\theta,\lambda_2 \right].\nonumber\\&&\label{chi1}\end{eqnarray}
Equation \ (\ref{chi1}) is the central analytic result of the 
present work \cite{PTH03}. It applies to arbitrary states: mixed or pure
(Gaussian or non-Gaussian)
two-mode resource state  $AB$ and classical 
or nonclassical input state $in$.

 We multiply Eq.\ (\ref{chi1})
by $\exp{(|\lambda_2|^2/2)}$ and get
  \begin{eqnarray}\chi_{B,out}^{(N)}(\lambda_2)&=& \chi_{in}^{(N)}(\lambda_2)
\nonumber\\&&\times
\chi_{AB}\left[\Re e(\lambda_2)\cot{\theta}-i \Im m ( \lambda_2) 
\tan\theta,\lambda_2 \right],\nonumber\\&&\label{chi2}\end{eqnarray}
where the superscript $(N)$ stands for normally ordered.

In his seminal
 paper on the coherent states of the electromagnetic
field, Glauber defined the density operator of the superposition of
two fields \cite{G63}:
\begin{equation}   \rho_{S}=\int{\rm d}^{2}\beta P_{2}(\beta)D(\beta)\rho_{1}
D^{\dag}(\beta).  \label{2.1}\end{equation}
In Eq.\ (\ref{2.1}), the field (2) is classical having the $P_{2}$-
representation and $\rho_{1}$ is the density operator of the field (1).
By considering the case in which the field (1) is also classical and 
described by the $P_{1}$-representation,
 Glauber obtained the well-known convolution law for a
classical superposition: 
\begin{eqnarray}  P_{S}(\beta)&=&
\int{\rm d}^{2}{\beta_{1}} P_{1}(\beta_{1}) P_{2}(\beta-\beta_{1}) 
\nonumber\\&&
=\int{\rm d}^{2}{\beta_{2}} P_{2}(\beta_{2}) P_{1}(\beta-\beta_{2}).
 \label{2.2}\end{eqnarray}
 
Subsequently, Eq.\ (\ref{2.1}) was  the  starting point in studying  the
case when the field (1) is allowed to be in a nonclassical state 
\cite{kim,PT96,PT97}. It was shown that the multiplication
 law of the normally ordered CF's,
\begin{equation}
\chi_{S}^{(N)}(\lambda)=\chi^{(N)}_{1}(\lambda)
\chi^{(N)}_{2}(\lambda), \label{2.7}\end{equation}
 derived from Eq\ (\ref{2.1}) is valid only when at least one of the two fields is
classical. Actually, in  Ref.\cite{kim} Kim and Imoto
showed that Eq.\ (\ref{2.7}) written for two {\it nonclassical} fields
leads to unphysical results.

Therefore our Eq.\ (\ref{chi2}) describes the superposition between
the input field $in$ and a one-mode field, say $M$,  extracted 
from the entangled 
$AB$ field 
by the measurement performed by Alice followed by the 
phase-space translation 
 performed by Bob. According to Eq.\ (\ref{2.7}) the function 
$\chi_{AB}\left[\Re e(\lambda_2)\cot{\theta}-i \Im m ( \lambda_2) 
\tan\theta,\lambda_2 \right]$ is its normally-ordered CF. Denote its usual CF as
\begin{eqnarray}\chi_{M}(\lambda_2)&=&\exp{(-|\lambda_2|^2/2)}
\nonumber\\&&\times
\chi_{AB}\left[\Re e(\lambda_2)\cot{\theta}-i \Im m ( \lambda_2)\tan {\theta},
\lambda_2\right].
\nonumber\\&&
\label{chiM}\end{eqnarray}

Equation \ (\ref{chiM}) tells us that the properties of 
the field state $M$ are fully determined by the  two-mode 
resource state. In the following we focuse on this issue.

\section{Properties of the state $M$}
Most of the work on continuous-variable teleportation protocol
 was done for two-mode Gaussian resources.
This is highly motivated by the experimental possibilities 
of producing entangled states. 
Originally, the quality of the teleportation protocol was quantified
 by the overlap of the states $in$ and 
$out$ for pure states \cite{BK,BL}, or the Uhlmann fidelity 
for mixed Gaussian states \cite{PTH03,Ban2}. So defined, 
the {\em fidelity of teleportation} depends on the input state. 
Equation \ (\ref{chi1}) suggests an alternative way of evaluating the 
accuracy of the teleportation process by means of the properties of the  CF, 
Eq.\ (\ref{chiM}). In the following we consider an arbitrary input state
 while an undisplaced two-mode Gaussian state is shared by the sender and 
 receiver. Recall that the CF
of an undisplaced two-mode Gaussian state is entirely specified by its covariance matrix ${\cal V}$ 

\begin{equation}
\chi_G(x)=\exp{\left(-\frac{1}{2}x^T {\cal V} x \right)},
\label{CF} 
\end{equation}
In Eq. \ (\ref{CF}) $x^T$ is a real row vector $(x_1\; x_2\; x_3\; x_4)$. 
The real, symmetric, and positive $4\times 4$ covariance matrix 
${\cal V}$ has the following block structure:
\begin{eqnarray}
{\cal V}=\left(\begin{array}{cc}{\cal V}_1&{\cal C}\\
{\cal C}^T&{\cal V}_2 \end{array}\right),
\label{cftm} 
\end{eqnarray}
where ${\cal V}_1$, ${\cal V}_2$, and ${\cal C}$ are $2\times 2$
matrices. Their entries are correlations of the canonical operators
$q_j=(a_j+a_j^{\dag})/{\sqrt{2}},\; p_j=(a_j-a_j^{\dag})/(\sqrt{2}i)$,
where $a_j $ and $a_j^{\dag}$,  $(j=1,2)$, are the amplitude operators 
of the modes. ${\cal V}_1$ and ${\cal V}_2$ denote the symmetric 
covariance matrices for the individual reduced one-mode squeezed thermal
 states, while the matrix ${\cal C}$ contains the cross-correlations 
between modes. Equivalently, the CF can be written as the complex-valued
 Gaussian function
\begin{eqnarray}
&&\chi(\lambda_1,\lambda_2)=\exp{[-(A_1+\frac{1}{2})|\lambda_1|^2
-\frac{1}{2}B_1^*
\lambda_1^2-\frac{1}{2}B_1(\lambda_1^*)^2]}
\nonumber \\ &&\times
\exp{[-(A_2+\frac{1}{2})|\lambda_2|^2-\frac{1}{2}B_2^*
\lambda_2^2-\frac{1}{2}B_2(\lambda_2^*)^2]}
\nonumber \\ &&\times
\exp{[- F\lambda_1^*\lambda_2-
 F^*\lambda_1\lambda_2^*
+G\lambda_1^*\lambda_2^*+G^*\lambda_1\lambda_2
]},\nonumber\\&&\label{cfcv}
\end{eqnarray}
leading to useful expressions of the entries of the covariance matrix 
${\cal V} $ in terms of the parameters $A_1,A_2,B_1,B_2, F, G$ 
\cite{PTH01}. After going through this correspondence we have
 finally obtained the $2\times2$ covariance matrix of the one-mode
 Gaussian state $M$
via Eq.\ (\ref{chiM}),
\begin{eqnarray}
{\cal V}_M=\left(\begin{array}{cc}\sigma(qq)&\sigma(qp)\\
\sigma(qp)&\sigma(pp)\end{array}\right).
\label{vm}\end{eqnarray} 
The entries of the covariance matrix ${\cal V}_M$ are 
\begin{eqnarray}\sigma(qq)&=&\frac{1}{2}+\sigma(q_2q_2)+
(\tan \theta)^2 \sigma(q_1q_1)- 2 \tan \theta \sigma(q_1q_2)\nonumber\\&&
\label{qq}\end{eqnarray}
\begin{eqnarray}\sigma(qp)
&=&\sigma(q_2p_2)- \sigma(q_1p_1)
+\cot \theta \sigma(q_2p_1)-
\tan \theta \sigma(q_1p_2)\nonumber\\&&\label{qp}\end{eqnarray}
\begin{eqnarray}\sigma(pp)&=&\frac{1}{2}+\sigma(p_2p_2)+
(\cot\theta)^2 \sigma(p_1p_1)+2 \cot \theta \sigma(p_1p_2).
\nonumber\\&&\label{pp}
\end{eqnarray}
It is interesting to rewrite Eqs.\ (\ref{qq})--\ (\ref{pp}) by introducing two 
commuting operators as combinations of the canonical operators of the 
two-mode resource field
\begin{equation} \hat Q:=\cos \theta \hat q_{2}-\sin \theta \hat q_1,\;\;
 \hat P:=\sin \theta\hat p_2+\cos \theta \hat p_1. \label{b2} \end{equation}
The operators $\hat Q, \hat P$ are closely related
 to those measured by Alice, Eq.\ (\ref{b1}). We have
\begin{eqnarray}&& \hat Q (\hat q_1, \hat q_2)=\hat q (\hat q_{A}\rightarrow \hat q_1, \hat q_{in}\rightarrow \hat q_2)\nonumber\\
&&\hat P(\hat p_1, \hat p_2)=\hat p(\hat p_A\rightarrow \hat p_1, \hat p_{in}\rightarrow \hat p_2)\label{b3}\end{eqnarray}
Equations \ (\ref{qq})--\ (\ref{pp}) become
\begin{eqnarray}\sigma(qq)=\frac{1}{2}+\frac{1}{(\cos \theta)^2}\langle 
\hat Q^2\rangle
\label{qq1}
\end{eqnarray}
\begin{eqnarray}\sigma(qp)=\frac{2}{\sin (2\theta)}\langle 
\hat Q \hat P\rangle\label{qp1}\end{eqnarray}
\begin{eqnarray}\sigma(pp)=\frac{1}{2}+\frac{1}{(\sin \theta)^2}\langle 
\hat P^2\rangle.\label{pp1}
\end{eqnarray}
The expressions \ (\ref{qq1})--\ (\ref{pp1}) allow us to enforce 
the interpretation given to the function \ (\ref{chiM}) by checking on the 
Robertson-Schr\"odinger uncertainty relation
\begin{eqnarray}\det{\cal V}_M=\sigma(qq)\sigma(pp)- 
(\sigma(qp))^2 \geq \frac{1}{4}.\label{UR}\end{eqnarray}
As we always have
\begin{eqnarray}\langle 
\hat Q^2\rangle\langle \hat P^2\rangle-\langle 
\hat Q \hat P\rangle^2\geq 0,\label{UR1}\end{eqnarray}
the uncertainty relation \ (\ref{UR}) is verified and hence the
 measurement-induced $M$-state is a physical state.
Moreover, from Eqs.\ (\ref{qq1})--\ (\ref{pp1}) we can readily find that
 \begin{eqnarray} {\cal V}_M-\frac{1}{2}I_2 \geq 0, \label{cl}\end{eqnarray}
where $I_2$ is the $2\times 2$ identity matrix.
The semipositiveness condition \ (\ref{cl}) defines the {\em classicality}
 of the one-mode Gaussian state $M$. This result is 
independent of the properties of the two-mode resourse state such as
 entanglement or $EPR$-correlations, being a consequence of the reduction of 
the three-mode quantum state by the measurement process. 

\section{The quality of the teleportation process}
Following Refs.\cite{HK,F,MF} we evaluate the teleportation quality in terms 
of the mean occupancy in the remote field $M$ which can be seen as 
the amount of noise distorting the properties of the input field state. We have
\begin{equation}N_{added}=\frac{1}{2}\left[\sigma(qq)+\sigma(pp)-1\right],
\label{ad}\end{equation}
and further
\begin{equation}N_{added}=\frac{1}{2}\left[\frac{1}{(\cos \theta)^2}\langle 
\hat Q^2\rangle+\frac{1}{(\sin \theta)^2}\langle 
\hat P^2\rangle\right].
\label{ad1}\end{equation}
From the equivalent expression 
\begin{eqnarray}N_{added}&=&\frac{1}{2}\left[\sigma(q_2q_2)+
(\tan \theta)^2 \sigma(q_1q_1)+\sigma(p_2p_2)\right. \nonumber\\&&\left.+
(\cot\theta)^2 \sigma(p_1p_1)+2 \cot \theta \sigma(p_1p_2)\right.
\nonumber\\&&\left.-
2\tan \theta \sigma(q_1q_2)\right], 
\label{ad2}\end{eqnarray}
we learn that the {\em unbalanced} measurement acts like
 a supplementary local squeezing of the mode 1. We can write
\begin{equation}(\tan \theta)\hat{q}_1\longrightarrow
 \hat{q^{\prime}}_1,\;\;(\cot\theta)\hat{p}_1\longrightarrow \hat{p^{\prime}}_1
\label{sub}\end{equation}
and reduce the discussion to the balanced beam splitter-case. It is 
thus sufficient to consider the balanced case when studying 
 the optimization possibilities of the teleportation process. 
According to Eq.\ (\ref{ad2}) the added noise is determined by the 
correlation properties of the two-mode state $AB$.
We denote by $b_1,b_2,c,d$ the parameters
in the standard form I of the covariance matrix of
 a class of two-mode Gaussian states locally related
 by squeezing operations and having thus the same entanglement \cite{Duan}.
 We have to minimize  the added noise, Eq.\ (\ref{ad2}) written for
 $\theta=\pi/4$ and under local squeezing operations denoted 
by $u_1$ and $u_2$ performed
on the standard  resource state $\rho^{(0)}_{AB}$ \cite{d1}. We write
 the two-variable function
\begin{eqnarray}N(u_1, u_2)&:=&\frac{1}{2}\left[b_1\left(u_1
+\frac{1}{u_1}\right)+b_2\left(u_2
+\frac{1}{u_2}\right)\right. \nonumber\\&&\left.-2\left(c\sqrt{u_1u_2}
+ \frac{|d|}{\sqrt{u_1u_2}}\right)\right].\label{f}
\end{eqnarray}
Minimization of Eq.\ (\ref{f}) with 
respect to $u_1$ and $u_2$ leads to an algebraic 
system whose solution is denoted by  $\vec{v}:v_1,v_2$: 
\begin{eqnarray}\frac{{b_1}({v_1}^2-1)}{v_1}=\frac{{b_2}({v_2}^2-1)}{v_2}
\label{f4a}\end{eqnarray}
\begin{eqnarray}b_1b_2(v_1^2-1)(v_2^2-1)=(c v_1v_2-|d|)^2.
\label{f4b}\end{eqnarray}
The general solution of this system is  complicated, 
arising finally from a fourth-order one-variable algebraic equation. Note that
Eq.\ (\ref{f4b})  coincides with one of the equations 
defining the standard form II of the covariance matrix  in Ref.\cite{Duan},
while Eq.\ (\ref{f4a}) is different from the  corresponding one first written 
 in Ref.\cite{Duan}. Now we have 
to recall that the standard
 form II of the covariance matrix was used by Duan {\em et al.} to give an
 separability criterion for two-mode Gaussian states based on the minimization
of the correlations of two ingeniously defined EPR-like  operators 
$P(\pm 1,\alpha),Q(\alpha)$. Generally,
 these  are {\em non-commuting} operators with the notable exception of 
the symmetric 
states for which they are proportional to our operators $P$ and $Q$, 
Eqs.\ (\ref{b2}) for $\theta=\pi/4$. Note that symmetric Gaussian states,
 $b_1=b_2$, have equal marginal purities. The algebraic system 
\ (\ref{f4a})--\ (\ref{f4b}) has the unique solution
\begin{equation}v_1=v_2=\sqrt{\frac{b_1-|d|}{b_1-c}}\label{u12}.\end{equation}
The squeeze factor \ (\ref{u12}) determines also the standard 
form II for which the minimal EPR correlations in the symmetric case are \cite{PT}
\begin{equation} \Delta_{EPR}=2[(b_1-c)v_1+(b_1-|d|)/v_1]=4\tilde{c}_{-}.\label{c1}
\end{equation}
Here $\tilde{c}_{-}:=\sqrt{(b_1-|d|)(b_1-c)}$ is the smallest
 symplectic eigenvalue of the partially transpose covariance
 matrix of the symmetric state. According to the inseparability 
criterion formulated in Ref.\cite{Duan}, a symmetric state is entangled 
iff $\Delta_{EPR}<2$ namely
$\tilde{c}_{-}<\frac{1}{2}$. This induces a condition to the minimal added 
noise obtained with the solution \ (\ref{u12}) for a symmetric entangled state
 \begin{equation} N_{min}(v_1,v_1)=2\sqrt{(b_1-|d|)(b_1-c)}< 1.
\label{c}\end{equation}

Therefore, we can see that an entangled Gaussian state used as a resource
 state in teleportation provides less 
noise in the output state. However, only for symmetric states a 
direct  relation between the amount of entanglement \cite{Giedke} of the resource 
state  and the quality of the teleportation could be established. 

To conclude, it appears that  a correspondence between the optimal added 
noise and the  entanglement of the two-mode resource state is unlikely 
to be found in general. This happens  because, according to the  criterion 
in Ref.\cite{Duan}, the inseparability is determined by the correlations 
of EPR-like defined operators
 which are non-commuting, while the teleportation fidelity
 depends on correlations of typical (commuting) EPR operators. Interestingly, 
we have found that in the  symmetric case, the minimal added 
noise is realized for  
states having the covariance matrix in standard form II \cite{Duan}.
\section*{Acknowledgments}
This work was supported by the Romanian 
CNCSIS through the grant 263/2004 for the University of Bucharest.
 

\begin{thebibliography}{99}
\bibitem{L}U. Leonhardt, Phys. Rev. A {\bf 48}, 3265 (1993). 
See Eq. (39). It can be directly checked that Eq. \ (\ref{a10})
is a commun eigenfunction for the commuting operators \ (\ref{op}).
\bibitem{G63}R. J. Glauber, Phys. Rev. {\bf 131},2766 (1963).
Note also that when the field (2) is in a $P$-representable
 Gaussian state, Eq.\ (\ref{2.1}) is the much recently studied 
Gaussian noise mapping.
\bibitem{d1} The standard state  $\rho^{(0)}_{AB}$ is defined as the 
two-mode Gaussian state whose covariance matrix is in the standard 
form I introduced in Ref.\cite{Duan};
 its one-mode reductions are thermal states.
\end{thebibliography}
\end{document}